\documentclass[fleqn,usenatbib,useAMS]{mnras}

\usepackage{newtxtext,newtxmath}

\usepackage[T1]{fontenc}
\usepackage{ae,aecompl}

\usepackage{graphicx}	
\usepackage[caption=false]{subfig} 
\usepackage{amsmath}	
\usepackage[dvipsnames]{xcolor}
\usepackage{subfig,graphicx}
\usepackage{enumitem}

\usepackage{hyperref} 
\hypersetup{
    colorlinks=true,
    linkcolor=blue,
    filecolor=magenta,      
    urlcolor=blue,
}

\newcommand{\be}{\begin{equation}}      
\newcommand{\ee}{\end{equation}}      
      
\newcommand{\bef}{\begin{figure}}      
\newcommand{\eef}{\end{figure}}      
\newcommand{\bea}{\begin{eqnarray}}    
\newcommand{\eea}{\end{eqnarray}}

\newcommand{\rf}{\texttt{rf }}

\newcommand{\arwv}{\texttt{ARWV }}

\title[Orbital precession of stars in the Galactic center]{Orbital precession of stars in the Galactic center}

\author[R. Capuzzo--Dolcetta and M. Sadun--Bordoni]{
R. Capuzzo--Dolcetta$^{1}$\thanks{E-mail: roberto.Capuzzodolcetta@uniroma1.it}
and M. Sadun--Bordoni$^{1,2}$
\\
$^{1}$Dep. of Physics, Sapienza, Univ. of Rome, P.le A. Moro 5, Rome, Italy\\
$^{2}$Max-Planck-Institut f\" ur Extraterrestrische Physik,
Giessenbachstrasse 1, 85748 Garching, Germany\\
}

\date{Accepted XXX. Received YYY; in original form ZZZ}

\begin{document}
\label{firstpage}
\pagerange{\pageref{firstpage}--\pageref{lastpage}}
\maketitle\noindent



\date{\today}

\begin{abstract}
The region around the center of our Galaxy is very dense of stars. The kinematics of inner moving stars in the Galaxy (the so called S-stars) has been deeply studied by different research groups leading to the conclusion of the existence of a very compact object (Sgr A$^*$, likely a supermassive black hole) responsible for their high speed. Here we start from the observational evidence of orbital apsidal line precession for the S2 (also called S0-2) star to investigate on a theoretical side what level of quality in such regime of relatively strong gravitational field is reached in the orbit angular precession determination when using a direct orbital integration of the star motion subjected to an acceleration computed in the post-Newtonian (PN) scheme up to different orders. This approach, although approximated and limited to particle speed not exceeding $\sim \ 0.3 c$,  allows the inclusion of various effects, like that of a possible spin of the central massive object. Our results show that the inclusion of PN terms above the standard 1PN term (the one corresponding to the classic Einstein-Schwarzschild estimate of pericenter advance) is compulsory to determine angular precession at sufficient level of accuracy for those penetrating stars that would allow to pick contemporary the value of the mass and of the spin of a rotating (Kerr-like) super massive black hole (SMBH).
We discuss how future observational data, together with a proper modelization, could allow the determination of both mass and spin of the SMBH of our Galaxy.
\end{abstract}



\section{Introduction}
The quantity of angular precession of the perihelion of Mercury is considered as one of the most striking confirmations of the validity of the Einstein's general relativity theory. As a matter of fact, also keeping into careful account the Newtonian effects induced on the Mercury's perihelion position by relevant planets (Venus, Earth, Mars, Jupiter and Saturn) the global effect underestimates for $42.98 \pm 0.05$ arcsec per century the modernly derived value of the Mercury's precession. It was Einstein \citep{ein1915a,ein1916} to determine in an approximated way the pericenter advance of the orbit expected in a Schwarzschild geodesic.

Let us note that the other possible causes of an increment in the Mercury's precession additional to the planetary one could be,  in the Newtonian framework,  given by the Sun potential quadrupole contribution, represented by the standard $J_2$ moment. The induced angular precession is linear in $J_2$ and modern values of $J_2$ gives an amount of precession up to only $0.07\%$ of the cited general relativistic 1st order effect \citep{par2017}. Also the deviation from the Newtonian inverse square dependence on the distance can be ruled out because, in spite of the very modest modification to the Newtonian interaction force needed to justify the observed Mercury's perihelion advance (a power of $-2.00000016$ instead of Newtonian $-2$ was suggested by \cite{hal1894} and further identified as $-2.0000001574$ by \cite{new1895}, this modification would lead to an advance of the lunar perigee incompatible with modern data on the lunar orbit. 

In recent times, the evidence of the existence of a very massive compact object in the Milky Way center (in the position of the Sgr A$^*$ radio source) provides a beautiful chance to study the effects of an intermediate-strong gravitational field on the kinematics of surrounding objects, thing so far limited to the relatively weak Sun's gravitational field.

In this paper we investigate the angular precession of the apsidal line in the Schwarzschild's bound geodesic, aiming at determining at what level of gravitational field strength around a massive object of the size of Sgr A$^*$ the deviation from the lowest order estimate of the  precession is appreciable. This analysis has a strong astrophysical motivation because of the ever improving quality of astrometric observations of stars around the Galactic center by the groups lead by the 2020 Nobel laureates R. Genzel at the MPE in Garching and A. Ghez at UCLA in Los Angeles. These stars are either called Sn (with n an identifying integer) by the Genzel's group or S0-l by the Ghez's group (increasing integer l corresponds to increasing distances from Sgr A$^*$).
\\
In particular, after the pericenter passage in 2018 of the S2 star (also denominated as S0-2 in the Ghez's group papers), both teams were able to detect both a gravitational redshift of spectral lines as predicted by General Relativity in regime of strong gravitational fields and the transverse Doppler effect as predicted by Special Relativity in high speed regime \citep{abu18}, that together give a total contribution to the redshift of $z\approx 200 \, \frac{km \, s^{-1}}{c}$.
From data up to the end of 2019, the group lead by R. Genzel was able to robustly detect also the in-plane, prograde  precession of its pericenter \citep{abu20}, that was claimed to be compatible with the 1PN classic Einstein-Schwarzschild estimate, and is equal to $\delta \varphi \sim$ $12.1^\prime$ over a radial period.

Starting from these observational results, in sect. 2 we study the apsidal line precession of the orbit of S2 and of a set of hypothetical, more penetrating, stars around Sgr A*, that we consider as a a Schwarzschild SMBH. This with the aim of comparing the best direct computation possible of this shift around the Schwarzschild singularity with what can be obtained integrating numerically the orbital motion using post-Newtonian terms in the evaluation of the acceleration. We examined also the validity of different 'analytical'  approximations of the apsidal line precession available in the literature. In sect.  3 we discuss the role of the spin of the SMBH in the angular precession and determine the orbital characteristics of more penetrating, potentially observable, S-stars that could allow to measure both the mass and the spin of Sgr A*. Finally, in sect.  4 we draw some conclusions and outline perspectives. 


\section{Classical and relativistic orbital precession in spherical symmetry}
\subsection{Classical precession}
The so called Bertrand's theorem \citep{ber73} states that only two types of central-force deriving by spherical scalar potentials $U(r)$
\footnote{we adopt the sign convention on the potential such that the force per unit mass is $\textbf{F} = \nabla U$.}
show the property that all bound trajectories are also \textit{closed} trajectories. They are the (attractive) (i) point mass potential ($U \propto 1/r$),  and the (ii) homogeneous sphere harmonic potential ($U\propto -r^2$).
In both cases, bound trajectories are ellipses, with the difference that in (i) the force center is in one of the foci while in (ii) it is in the ellipse center. In all other spherical potentials which admit bound trajectories the trajectories of the test particle are not closed.
Thus, in spherical potentials of gravitational origin ($dU/dr<0$)
orbits precede within two circumferences of inner radius $r_p$ (pericenter distance) and $r_a$ (apocenter distance) in a way such that the angular position of the apsidal line on the orbital plane
shifts along a complete radial oscillation (from $r_p$ to $r_a$ and back) by an amount \citep{rcdbook19}

\begin{equation}
\Delta\varphi=  2L \int\limits_{r_p}^{r_a} \frac{dr}{r^2\sqrt{2\left[E-\dfrac{1}{2}\dfrac{L^2}{r^2}+U(r)\right]}} 
\end{equation}

where $E$ and $L$ are the specific (per unit mass) orbital energy and absolute value of the angular momentum of the particle in motion.
Obviously, both $r_p$ and $r_a$ are functions of $E$ and $L$.
Given that for a closed orbit $\Delta\varphi =2\pi$, the angular shift of the apsidal line per full radial oscillation of a precessing orbit is $\delta \varphi= \Delta\varphi -2\pi$, so that (under the assumption of particle moving on its orbital plane in a prograde (counterclockwise) motion) $\delta\varphi >0$ corresponds to a \textit{prograde} or \textit{forward} precession (pericenter advance) and $\delta\varphi<0$ to a \textit{retrograde} or \textit{backward} precession.

The classical precession is usually considered as retrograde, i.e. in the opposite direction of the particle motion. The explanation is easy: in a Keplerian  force field (the one generated by a point-mass M, so that the density is formally represented by a Dirac's delta function centered in $r=0$, $\delta(r)$)
there is no precession, so, whenever the same mass $M$ is spread with radial symmetry over space with any regular density law $\rho(r)>0$ broader than the delta function, the absolute value of the force at any $r$ is smaller than in the point-mass case because 

\begin{equation}
M_\rho(r)=4\pi \int\limits_0^r\rho(r)r^2dr < M,
\end{equation}

for every $r$, so that in the case of the regular density distribution the gravitational acceleration (which measure the local curvature of the trajectory) is reduced. 

\subsection{Relativistic precession}

In the case of a massive, non-rotating (zero spin) black hole  the motion of particles of negligible mass (with respect to the BH)  around the singularity are described by the Schwarzschild geodesics \citep{sch16}.

In the Schwarzschild geodesics the angular shift of the apsidal line can be determined via an integral expression.  Assuming $r$ and $\varphi$ polar coordinates on the plane of motion (assumed as the $\theta =\pi/2$ plane), the differential equation which determines the trajectory of motion around the Schwarzschild's singularity is 

\begin{equation}
\left(\frac{dr}{d\varphi}\right)^2=\frac{r^4}{b^2}-\left(1-\frac{r_S}{r}\right)\left(\frac{r^4}{a^2}+r^2\right),
\label{eq:traj1}
\end{equation}
where $r_S$ is the Schwarzschild's radius, $r_S=(2Gm_\bullet)/c^2$, with $m_\bullet$ mass of the black hole and $c$ speed of light in vacuum, and $a$ and $b$ are two length scales defined as $a=L/c$ and $b=cL/E$, with $L$ and $E$ the previously introduced specific (per unit mass) orbital angular momentum and energy of the test particle of mass $m\ll m_\bullet$. Defining $u\equiv 1/r$, Eq. \ref{eq:traj1} becomes, for non radial orbits ($L \neq 0$ and so $r\neq 0$)

\begin{equation}
\begin{split}
\left(\frac{du}{d\varphi}\right)^2= &\frac{1}{b^2}-\left(1-ur_S\right)\left(\frac{1}{a^2}+u^2\right)=\\
= & r_Su^3 - u^2 +\frac{r_s}{a^2}u+ \frac{1}{b^2}-\frac{1}{a^2}.
\end{split}
\label{eq:traj2}
\end{equation}

The solutions of the cubic equation 

\begin{equation}
\label{eq:cubic}
 \left(\frac{du}{d\varphi}\right)^2= r_Su^3 - u^2 +\frac{r_s}{a^2}u+ \frac{1}{b^2}-\frac{1}{a^2}=0, 
\end{equation}

give the radial distances of stationarity of the $u(\varphi)=1/r(\varphi)$ relation (the trajectory). For initial conditions such that the three roots $u_i, i=1,2,3$ of Eq. \ref{eq:cubic} are all real, and ordering them as $u_1<u_2<u_3$, it can be shown \citep{dam88} that the angular shift of the periapsis line over one full radial oscillation is given by the expression

\begin{equation}
\begin{split}
\delta \varphi =& \Delta\varphi - 2 \pi  =\dfrac{4}{\sqrt{r_S(u_3-u_1)}}\int\limits_0^{\pi/2}\dfrac{d\varphi}{\sqrt{1-k^2\sin^2\varphi}} - 2\pi = \\
& =\dfrac{4}{\sqrt{r_S(u_3-u_1)}}\int\limits_0^1\dfrac{dt}{\sqrt{(1-t^2)(1-k^2t^2)}} - 2 \pi,
\label{eq:angshift}
\end{split}
\end{equation}

where 

\begin{equation}
k^2=\dfrac{u_2-u_1}{u_3-u_1}\geq 0.
\end{equation}


The integrals in Eq. \ref{eq:angshift} are known as complete elliptic integrals of the first kind, usually referred to as 
$K(k)$. They must be numerically evaluated to obtain the $\delta\varphi$ value once that the roots $u_1,~u_2,~u_3$ of the cubic Eq. \ref{eq:cubic} have been obtained. The formal expression of the $u_i$ roots is available in trigonometric form (after reduction of the cubic to depressed form obtained by a substitution $t(u)$ that transforms the equation into one containing only a constant and $t^3$ and $t$ terms) but it is quite complicated and so we found more convenient choosing a numerical approach via bisection method followed by a Newton-Raphson refinement which lead to an accuracy of $18$ digits. 
After this, the function $K(k)$ has been numerically evaluated by mean of the subroutine \rf  distributed under the GNU LGPL license \citep{carl79,carlnot81}, which is the same used in the Wolfram Mathematica package. In particular, we used the Fortran 90 version provided by John Burkardt. In its use, we required a relative truncation error less than $10^{-20}$.
Following the described methodology, we could evaluate at enormous precision the `exact' angular precession per radial period in the case of stars of negligible mass around the Schwarzschild singularity of mass $m_\bullet$, covering a wide range of initial conditions starting from those of the S2 star. This with the aim to compare these results with those obtainable from direct integration of the star motion in the post-Newtonian force field at the various orders, as well as with those obtained from apsidal line shift approximations available in the literature.

\subsection{A comparison of `exact' periapsis precession with PN estimates}

It is well known that for planet Mercury the classical first order general relativistic estimate of its apsidal line precession as given by Einstein himself \citep{ein1915a,ein1916} was totally satisfactory to explain the unexplained extra angular perihelion advance of 42.98 arcseconds per century. The precision of the lowest order approximation is not surprise because even at its pericenter $r_{p,M}$, Mercury is subjected to a relatively weak gravitational field, as quantified by the ratio $r_{p,M}/r_{S,\odot}\simeq 1.5603 \times 10^4$, 
($r_{S,\odot}\simeq 2.9541$ km is the Sun Schwarzschild radius)
and its relatively low pericenter speed $v_{p,M}$, $v_{p,M}/c\simeq 1.8889\times 10^{-4}$. 
The situation for S2 is different because (see also Tab. 2) its pericenter to Sgr A$^*$ Schwarschild radius is less than $1/10$ smaller, $r_{p,S2}/r_{S,\bullet} \simeq 1.41421 \times 10^{3}$ and the speed ratio about 100 greater, $v_{p,S2}/c \simeq 0.02581$. In post-Newtonian treatment, both strength of gravitational field and object speed relatively to speed of light are consistently taken into account, so it is not a priori obvious that for stars like S2, and a fortiori for deeper plunging stars, the simplest 1st order approximation is enough to determine accurately its real relativistic pericenter advance.\\
Given this, we studied the precession of the orbit of the S2 star as coming from its data published in \cite{abu20} and of other  (hypothetical) more `penetrating' stars by direct integrations of their orbital motion using our own high quadruple-precision Fortran code in the assumption of negligible mass of the test star that leads to a one body problem.  We considered post-Newtonian terms in the computation of the acceleration of the test star. For the purposes of this paper, which involve high accuracy, we implemented a one-body code which integrates the equation of motion of the test star by mean of a high order Runge Kutta method with variable time step in a quadruple precision (128-bit FPA) environment. 
As first choice, we adopted the Runge Kutta 8(5,3) method due to \cite{dopr80}. We have made a careful check of results coming from this integrator with results obtained with the use of our own quadruple precision fortran 90 version of the regularized code \arwv. This code was originally developed by \cite{miktan99b,miktan99a} and later modified to include relativistic recoil velocity and external potential \citep{chacd19,chacd21}. We noted that results correct at the precision required for the scopes of this paper could be obtained by mean of a simple and fast Runge Kutta 4 integrator, provided a careful variable time stepping procedure is used or a constant time step of the order of $10^{-7}$ times the full orbital time extension (which we usually adopt as 8 radial periods). We have, also, carefully checked that integrations performed with quadruple precision were an excellent compromise between accuracy requirement and computing time.\\

In the general case of $N$ S-stars represented as point like objects identified by position vectors ${\mathbf{r}}_{i}$ and masses $m_i$ (for $i=1,2,...,N$) the equations of motion are the following

\begin{equation}
\label{eq_NBsystem}
    \ddot{\mathbf{r}}_{i} = G \sum_{\substack{j=1 \\ j\neq i}}^N m_j \frac{\mathbf{r}_{j} - \mathbf{r}_{i}}{|\mathbf{r}_{j} - \mathbf{r}_{i}|^3} + G m_\bullet \frac{\mathbf{r}_\bullet - \mathbf{r}_{i}}{|\mathbf{r}_\bullet - \mathbf{r}_{i}|^3} +
    \mathbf{f}_{\textrm{PN},i} +  \mathbf{f}_{ext,i},
\end{equation}

where the dots represent time differentiation,  $G$ is the Newton's gravitational constant and $\mathbf{r}_\bullet$ is the massive BH position vector (in the limit $\sum_j m_j/m_\bullet \rightarrow 0$, $\mathbf{r}_\bullet$ coincides with the star system center of mass (c.o.m.), i.e. $\mathbf{r}_\bullet \rightarrow 0$ in the center of mass frame). 
In Eq. \ref{eq_NBsystem},  $\mathbf{f}_{\textrm{PN},i}=\mathbf{f}_{\textrm{PN}}(\mathbf{r}_{i},\mathbf{v}_{i})$ is the post-Newtonian force per unit mass over the zero-th order Newtonian acting on the $i$th S-star as due to the presence of the massive BH, while $\mathbf{f}_{ext,i}=\mathbf{f}_{ext}(\mathbf{r}_{i})$ is the contribution of the rest of the stars
where the S-star cluster is embedded in and of possible diffuse matter.

It can be proven that in presence of a dominant (super massive) ``particle'' (the massive black hole, even better referred to as Compact Massive Object, CMO) the self-interaction among the S-stars is negligible. Moreover, we will not deal here with the ``external'' contribution (which as we said in the previous subsection gives a ``classical'' precession), so that Eq. \ref{eq_NBsystem} converts into a two-body PN problem that for $m_i\ll m_\bullet$ further reduces to a one-body problem. With the adoption of the origin of reference frame in the center of mass and a change of variables to relative (with respect to the massive BH) coordinates $\mathbf{r}=\mathbf{r}_i-\mathbf{r}_\bullet$, so that $\mathbf{n}=(\mathbf{r}_i-\mathbf{r}_\bullet)/r$ is the unit vector in the direction from the BH to the $i$th S-star and $\mathbf{v}=\mathbf{v}_i-\mathbf{v}_\bullet$ is the relative velocity, the differential equation of the relative motion of the $i$th generic test star reduces to

\begin{equation}
\label{eff_oneb_1}
\ddot{\mathbf{r}} = - G m_\bullet \frac{\mathbf{r}}{r^3} +
    \mathbf{f}_{\textrm{PN}}.   
\end{equation}

In the present case of non-spinning massive black hole, the PN contribution to the classic Newtonian acceleration of the $i$th star can be given in the form

\begin{equation}
\label{Eq:10}
 \mathbf{f}_{\textrm{PN}} = 
  \sum\limits_{k=1}^{k_{max}} c_k(r)\left[A_k (r,\dot r, v;\eta) \mathbf{n}
 + B_k(r,\dot r,v;\eta) \mathbf{v}\right],
\end{equation}

where $k$ is a positive integer or half-integer referring to the $k$PN order, up to the $k_{max}$ order, which we assume here $k_{max}=3.5$.
In Eq. \ref{Eq:10}:

\begin{equation}
c_k(r)=\frac{1}{r}\left(\frac{Gm_\bullet}{c^2r}\right)^k,
\end{equation}

and the functions $A_k$ and $B_k$ depend upon $r$, $\dot r$ and $v$ other than on the so called ``symmetric mass ratio'', $\eta$, defined as the ratio between the reduced mass, $\mu$, of the ($m_\bullet$, $m_i$) pair to the total pair mass 
\begin{equation}
\eta=\frac{\mu}{m_\bullet + m_i}=\frac{m_\bullet m_i}{(m_\bullet + m_i)^2}.   
\end{equation}
The general expressions of the $A_k$ and $B_k$ functions are easily obtained from, e.g., \cite{mowill04}.


We have to take into account that, in the Schwarzschild case, General Relativity does not produce $0.5$PN or $1.5$PN contributions to the metric or to the equations of motion and that the 1PN, 2PN and 3PN terms $(\mathbf{f}_{\mathrm{1PN}} , \mathbf{f}_{\mathrm{2PN}}~ \mathrm{and}~ \mathbf{f}_{\mathrm{3PN}})$ are non dissipative and responsible for the pericentre angular shift \citep{kupi2006,amaro2018}, 
while the 2.5PN term (radiation reaction term) is dissipative and accounts for energy loss via gravitational waves \citep{blanchet2014}. \\

The dependence on $\eta$ of the PN accelerations in our range of PN approximation is up to its third power \citep{mowill04}.
For a very massive black hole, it is $\eta \sim m_i/m_\bullet $, which is very small. In our case, given an estimated mass range $0.5 \lessapprox m_i/M_\odot \lessapprox 20$ for the S-stars (see for instance \cite{habibi2017,habibi19}), it results $1.19 \times 10^{-7} \leq \eta \leq 4.76\times 10^{-6}$. This means that at a first glance letting $\eta=0$ in the PN expressions of the acceleration (that can be taken from \cite{mowill04}) of the ``test'' star of mass $m_i$ in the $m_\bullet$ field would be accurate enough.
Actually, the expressions of $A_k$ and $B_k$ simplify significantly letting $\eta=0$ in their general expressions (see Tab. 1).  
Anyway, due to that we are looking for tiny effects (identification of the role of PN terms over 1st order in the periapsis angular precession) we will check in this paper the role of the non-zero value of the $\eta$ parameter. \\

\begin{table}
\label{tab:1}
\begin{center}
\caption{Post-Newtonian ``mass'' terms $A_k$ and $B_k$, whose index $k$ is in first column, in the limit $\eta=0$.}
{
\begin{tabular}{|c|c|c|}\hline 
$k$PN & {\bf $A_k$}  & {\bf $B_k$}  \\  \hline
0.5& 0 & 0\\
1& $-v^2+4\dfrac{Gm_\bullet}{r}$ & $4\dot r$ \\
1.5& 0 & 0\\
2& $-9\dfrac{Gm_\bullet}{r}+2\dot r^2$& $-2\dot r$ \\
2.5 &0& 0 \\
3& $16\dfrac{Gm_\bullet}{r}- \dot r^2$ & $4\dot r$ \\
3.5 & 0 & 0\\
\hline
\end{tabular}
}
\end{center}
\vspace{1mm}
\end{table}

\begin{table}
\label{tab:2}
\begin{center}
\caption{First set of circum-SMBH orbits chosen in this paper. Col. 1 gives the index referring to initial velocity angle with the positive $x$ axis, such that $\varphi_j=\pi/2+j\pi/20$; col. 2 and 3: initial velocity components; col. 4 and 5: Newtonian pericenter and apocenter distances (in units of SMBH Schwarzschild's radius); col. 6: $v/c$ at pericenter; col. 7: orbital eccentricity.}
{\scriptsize
\begin{tabular}{ |c|c|c|c|c|c|c| }
\hline 
$j$ & $\dot x_0$ & $\dot y_0$ & $r_p/r_S$ & $r_a/r_S$ & $v_p/c$ & $e$ \\
\hline
0  & 0 & $0.349792$ & $1.41421\times 10^3$ & $2.31059\times 10^4$ & $0.02581$ & $0.884649$\\
1  & $-0.0547195$ & $0.345486$ & $1.37741\times 10^3$ & $2.31427 \times 10^4$& $0.02618$  & $0.887651$ \\
2  & $-0.108092$ & $0.332672$ & $1.27130\times 10^3$ & $2.32488\times 10^4$& $0.02731$ & $0.896305$  \\
3  & $-0.158802$ & $0.311667$ & $1.10805\times 10^3$ & $2.34121\times 10^4$& $0.02936$ & $0.909621$\\
4  & $-0.205603$ & $0.282988$ & $9.05680\times 10^2$ & $2.36144\times 10^4$ & $0.03261$ & $0.926127$\\
5  & $-0.247340$ & $0.247340$ & $6.85487 \times 10^2$ & $2.38346\times 10^4$& $0.03766$  & $0.944088$\\
6  & $-0.282988$ & $0.205603$ & $4.69404\times 10^2$ & $2.4070\times 10^4$  & $0.04572$ & $0.961712$\\
6.5 & $-0.298247$ & $0.182766$ & $3.69311\times 10^2$ & $2.41455\times 10^4$ & $0.05164$  & $0.969871$\\
7  & $-0.311667$ & $0.158802$ & $2.77816\times 10^2$ & $2.42423\times 10^4$ & $0.05966$  & $0.977340$ \\
7.5 & $-0.323166$ & $0.133860$ & $1.96673\times 10^2$ & $2.43181\times 10^4$ & $0.07102$ & $0.983953$\\
8  & $-0.332672$ & $0.108092$ & $1.27924\times 10^2$ & $2.43922\times 10^4$ & $0.08819$  & $0.989566$\\
9  & $-0.345485$ & $0.0547195$ & $32.6556$ & $2.44875\times 10^4$ & $0.1749$  & $0.997336$\\
\hline
\end{tabular}
}
\end{center}
\vspace{1mm}
\end{table}

\begin{table}
\label{tab:3}
\begin{center}
\caption{For the cases of Tab. 2: col. 1 is the angle index; col. 2 gives the ``exact'' periapsis angular shift per radial period (Eq. \ref{eq:angshift}); cols. 3, 4 and 5 are, respectively, the differences between the ``exact'' $\delta\varphi$ and those obtained by orbital integrations accounting for PN contribution up to 1st, 2nd and 3rd order respectively, $(\delta \varphi)_{kPN}$, according to the definition $(\delta^2 \varphi)_{kPN}=\delta\varphi-(\delta\varphi)_{kPN}$, $k=1,2,3$. All the values are in arcmin.}
{\scriptsize
\begin{tabular}{ |c|c|c|c|c|c|c|c|c| }
\hline 
$j$ &  $\delta\varphi$ & $(\delta^2 \varphi)_{\mathrm{1PN}}$ &  $(\delta^2 \varphi)_{\mathrm{2PN}}$  &  $(\delta^2 \varphi)_{\mathrm{3PN}}$\\
\hline
$0$ &   $12.1750$ & $ 6.24847\times 10^{-3}$  &   $1.03474\times 10^{-3}  $&   $1.03760\times 10^{-3} $\\
$1$ &  $12.4810$  & $ 6.56605\times 10^{-3}$  & $  9.42230\times 10^{-4}$  & $  9.46045\times 10^{-4}$ \\
$2$ &  $13.4628$ & $ 7.48634\times 10^{-3}$  & $  1.11675\times 10^{-3}$  & $  1.12152\times 10^{-3}$ \\
$3$ & $15.3426$ & $ 9.76467\times 10^{-3}$  & $  1.50013\times 10^{-3}$ & $   1.50681\times 10^{-3}$ \\
$4$ & $18.6183$  & $ 1.39523\times 10^{-2} $& $   1.79100\times 10^{-3}$ & $   1.80435\times 10^{-3} $\\
$5 $& $24.3909$ & $ 2.35023\times 10^{-2} $& $   2.63405\times 10^{-3} $ & $  2.66266\times 10^{-3} $\\
$6 $& $35.3511$ & $ 4.79546\times 10^{-2}  $& $  4.06647\times 10^{-3}  $ & $  4.15039\times 10^{-3} $\\
$6.5 $& $44.7945$ & $ 7.57828\times 10^{-2}  $& $ 5.19180\times 10^{-3}  $& $ 5.36728\times 10^{-3} $\\
$7 $& $59.4513$ & $  0.131653  $& $  6.91986\times 10^{-3}  $& $ 7.33566\times 10^{-3} $\\
$7.5 $& $83.9465$ & $ 0.246902  $& $   -3.25775\times 10^{-3}  $& $ -2.08282\times 10^{-3} $\\
$8$ & $129.531$ & $ 0.614487$ & $  1.25275\times 10^{-2}  $& $ 1.68762\times 10^{-2} $\\
$9 $& $533.138$ & $ 10.9925  $& $ -0.244080$ & $ 8.37402 \times 10^{-2}$\\

\hline
\end{tabular}
}
\end{center}
\vspace{1mm}
\end{table}

\begin{table}
\label{tab:4}
\begin{center}
\caption{Second set of circum-SMBH orbits chosen. The stars have all same eccentricity of S2 ($j=0$) but a progressively smaller semi major axis and radial period. \\
Col. 1: index of the star, with same pericenter distance than star with same $j$ in Tab. 2; 
col. 2: initial position $(x_0,0)$, where $x_0>0$ is the apocenter of the orbit; col. 3:  initial velocity $(0,\dot{y}_0)$;  col. 4 and 5: semi-major axis and orbital radial period; col. 6: $v/c$ at pericenter.
}
{\scriptsize
\begin{tabular}{ |c|c|c|c|c|c|c| }
\hline 
$j$ & $x_0$ & $\dot y_0$ & $a$ (mpc) & $T_r$ (yrs)& $v_p/c$ \\
\hline
0  &  $0.942762$& 0.349792&$4.99996$ & $16.0580$& 0.02581\\
1  & $0.918230$& 0.354434&$4.86985$ & $15.4354$& 0.02616\\
2  & $0.847494$& 0.368929& $4.49471$ & $13.6867$& 0.02723\\
3  &$0.738663$& 0.395173& $3.91752$ & $11.1370$& 0.02916\\
4  & $0.603758$& 0.437099&$3.20205$ & $8.23014$& 0.03226\\
5  &$0.456967$ & 0.502422& $2.42354$ & $5.41953$& 0.03708\\
6  & $0.312922$ & 0.607146&$1.65959$ & $3.07136$& 0.04481\\
6.5 & $0.246243$& 0.684429&$1.30596$ & $2.14416$& 0.05051\\
7  & $0.185201$& 0.789204& $0.982218$ & $1.39872$& 0.05824\\
7.5 &$0.131134$& 0.937891&  $0.695475$ & $0.833545$& 0.06922\\
8 &$0.085279$& 1.163029& $0.452278$ & $0.437307$& 0.08583\\
9 &$0.021769$ & 2.301909& $0.115454$& $0.056584$& 0.1699\\
\hline
\end{tabular}
}
\end{center}
\vspace{1mm}
\end{table}

\begin{table}
\label{tab:5}
\begin{center}
\caption{For the cases of Tab. 2: col. 1 is the angle index; cols. 2, 3 and 4 are, respectively, the differences between the ``exact'' periapsis angular shift per radial period (2nd column of Tab. 3) and the approximated estimates obtained at 1st, 2nd and 3rd PN order (see sect.  2.4), according to the definition $(\tilde{\delta}^2 \varphi)_{kPN}=\delta\varphi-(\tilde{\delta}\varphi)_{kPN}$, $k=1,2,3$. All the $\delta^2 \varphi$ values are in arcmin.}
{\scriptsize
\begin{tabular}{ |c|c|c|c|c|c|c|c|c|c| }
\hline 
$j$ &   $(\tilde{\delta}^2 \varphi)_{\mathrm{1PN}}$  &  $(\tilde{\delta}^2 \varphi)_{\mathrm{2PN}}$ & $(\tilde{\delta}^2 \varphi)_{\mathrm{3PN}}$\\
\hline
0 & $1.87719\times 10^{-2}$ & $-5.62586\times 10^{-4}$ & $-6.01322\times 10^{-4}$ \\
1 & $1.97729\times 10^{-2}$ & $-5.59533\times 10^{-4}$ & $-6.01321\times 10^{-4}$  \\
2 & $2.31560\times 10^{-2}$ & $-5.48654\times 10^{-4}$ & $-6.01313\times 10^{-4}$\\
3 & $3.03565\times 10^{-2}$ & $-5.22853\times 10^{-4}$ & $-6.01273\times 10^{-4}$ \\
4 & $4.51734\times 10^{-2}$ & $-4.59937\times 10^{-4}$ & $-6.01098\times 10^{-4}$\\
5 & $7.82970\times 10^{-2}$ & $-2.80643\times 10^{-4}$ & $-6.00294\times 10^{-4}$ \\
6 &0.165758 & $3.82428\times 10^{-4}$ & $-5.95587\times 10^{-4}$ \\
6.5 & 0.266862 & $1.40530\times 10^{-3}$ & $-5.85606\times 10^{-4}$ \\
7 & 0.470819 & $4.09467\times 10^{-3}$ & $-5.50857\times 10^{-4}$ \\
7.5 & 0.938767 & $1.25984\times 10^{-2}$ & $-3.97910\times 10^{-4}$ \\
8 &  2.22934 & $4.75675\times 10^{-2}$ & $5.47381\times 10^{-4}$ \\
9 & 36.3905 & 3.09585 & 0.290235\\
\hline
\end{tabular}
}
\end{center}
\vspace{1mm}
\end{table}

\begin{table}
\label{tab:6}
\begin{center}
\caption{As in Tab. 3,  but for the second set of orbits of Tab. 4.}
{\scriptsize
\begin{tabular}{ |c|c|c|c|c|c|c|c| }
\hline 
$j$ & $\delta\varphi$ & $(\delta^2 \varphi)_{\mathrm{1PN}}$ & $(\delta^2 \varphi)_{\mathrm{2PN}}$ &$(\delta^2 \varphi)_{\mathrm{3PN}}$\\
\hline
$0$ &   $12.1750$ & $ 6.24847\times 10^{-3}$  &   $1.03474\times 10^{-3}  $&   $1.03760\times 10^{-3} $\\
$1$ & $12.5009$  &  $6.40392\times 10^{-3} $ & $9.06944\times 10^{-4} $ & $9.10759\times 10^{-4} $ \\
$2$ & $13.5461$ &  $ 7.88307\times 10^{-3}$ & $1.42670\times 10^{-3} $ & $1.28555  \times 10^{-3} $ \\
$3$ & $15.5461$ &  $ 1.04771\times 10^{-2}$ & $1.96934\times 10^{-3}$ & $1.97697\times 10^{-3}$ \\
$4$  & $19.0286$  & $1.61896 \times 10^{-2}$ & $3.43704\times 10^{-3}$ & $3.44849\times 10^{-3}$ \\
$5 $ & $25.1614$ & $ 2.85778\times 10^{-2}$ & $6.24847\times 10^{-3}$ & $6.27899\times 10^{-3}$ \\ 
$6 $ & $36.8000$ & $6.18172\times 10^{-2}$ & $1.45836\times 10^{-2} $ & $ 1.40305\times 10^{-2}$  \\
$6.5 $ & $46.8264$ & $0.100300$ & $ 2.30751\times 10^{-2}$ & $ 2.32697\times 10^{-2} $ \\
$7 $ & $62.3868$ & $ 0.179028$ & $ 4.06761\times 10^{-2}$ & $ 4.25644\times 10^{-2}$ \\
$7.5 $ & $88.4088$ & $0.362251$ & $ 8.28934\times 10^{-2}$ & $ 8.42056\times 10^{-2}$\\
$8$ & $136.804$ & $ 0.872543 $ & $ 0.198471$ & $ 0.203064$ \\
$9 $ & $566.213$ & $15.7300$ & $3.13062$ & $ 3.49786$ \\

\hline
\end{tabular}
}
\end{center}
\vspace{1mm}
\end{table}

\begin{table}
\label{tab:7}
\begin{center}
\caption{For the set of orbits of Tab. 2: col. 1 is the angle index; col. 2 is the pericenter distance in mpc; cols. 3 and 4 are, respectively: (3) the difference between the ``exact'' periapsis angular shift per radial period in the zero-spin case (2nd column of Tab. 3) and that obtained accounting for PN contribution up to 3rd order including spin-orbit and quadrupole terms, namely $\delta^2 \varphi=\delta\varphi-(\delta\varphi)_{PN}$ (in arcmin); (4) the same quantity in col. 3, divided by the the ``exact'' periapsis angular shift per radial period in the zero-spin case.
We assumed $\chi =1$ and spin up, ${\bf s}_\bullet =(0,0,1)$ (cols. 3 and 4) and spin down, ${\bf s}_\bullet =(0,0,-1)$ (cols. 5 and 6).
}
{\scriptsize
\begin{tabular}{ |c|c|c|c|c|c|c|c| }
\hline 
$j$ & $r_p$ [mpc] & $(\delta^2 \varphi)_{\mathrm{PN}}$ &  $(\delta^2 \varphi)_{\mathrm{PN}}/\delta\varphi$ & $(\delta^2 \varphi)_{\mathrm{PN}}$ & 
$(\delta^2 \varphi)_{\mathrm{PN}}/\delta\varphi$\\
& &  $s_{\bullet z}=1$ & $s_{\bullet z}=1$ & $s_{\bullet z}=-1$ & $s_{\bullet z}=-1$  \\
\hline
$0$ & $0.576756$  & $0.223041$ & $1.83195\times 10^{-2}$ & $-0.223760$ & $-1.82965\times 10^{-2}$ \\
$1$ & $0.561748$  & $0.231358$ & $1.85362\times 10^{-2}$ & $-0.23135$ & $-1.85362\times 10^{-2}$ \\
$2$ & $0.518474$ & $0.259489$ & $1.92745\times 10^{-2}$ & $-0.259411$ & $-1.92687\times 10^{-2}$\\
$3$ & $0.451894$ & $0.315190$ & $2.05435\times 10^{-2}$ & $-0.315710$ & $-2.05774\times 10^{-2}$ \\
$4$ & $0.369363$ & $0.421259$ & $2.26261\times 10^{-2}$ & $-0.422640$ &  $-2.27003\times 10^{-2}$\\
$5$ & $0.279560$ & $0.631701$ & $2.58990\times 10^{-2}$ & $-0.635300$ & $-2.60466\times 10^{-2}$\\
$6$ & $0.191437$ & $1.10145$ & $3.11575\times 10^{-2}$ & $-1.11285$ & $-3.14800\times 10^{-2}$\\
$6.5$ & $0.150645$ & $1.57096$ & $3.50704\times 10^{-2}$ & $-1.59234$ &  $-3.55477\times 10^{-2}$\\
$7$ & $0.113301$ & $2.40308$ & $4.0421\times 10^{-2}$ & $-2.44542$ &  $-4.11332\times 10^{-2}$\\
$7.5$ & $0.0802245$ & $4.02508$ & $4.79481\times 10^{-2}$ & $-4.14552$ & $ -4.93289\times 10^{-2}$\\
$8$ & $0.0521712$  & $7.76170$ & $5.99217\times 10^{-2}$ & $-8.0154$ &  $-6.18803\times 10^{-2}$\\
$9$ & $0.0133179$ & $67.2602$ & $0.126159$ & $-73.6802$ & $- 0.138201$\\
\hline
\end{tabular}

}
\end{center}
\vspace{1mm}
\end{table}

For computational convenience, the units of measure are chosen such that $G=1$, as unit of mass we took the SMBH mass $m_\bullet =4.261\times 10^6$ M$_\odot$ \citep{abu20}
and as unit of length $D$ that subtended by the angle 250 mas at the distance, $R_0=8246.7$ pc \citep{abu20}, of the Sun to the Galactic center, that is $D=9.99528\times 10^{-3}$ pc. In such units, the speed of light in vacuum is $c=221.3988597$. \\

In Tab. 2 we give the initial velocity conditions for a set of orbits of same orbital energy of the S2 star (which, in this Table, corresponds to $j=0$) but a reducing orbital angular momentum $\mathbf{L}=L_z\mathbf{k}$, that corresponds to a progressive increase of the eccentricity and so to a reduction of the pericenter distance and to an increase of the apocenter distance. 
The data for the S2 star have been taken from Tab. E.1 of \cite{abu20}. Given the same orbital energy, all the orbits have the same semi-major axis and radial orbital period of S2, namely
\begin{equation}
    \begin{split}
        &a= 4.99996 \,\rm{mpc},\\
    &T_r=  16.0580 \, \rm{yr}.
    \end{split}
\end{equation}
The various rows in Tab. 2 are for various choices of the initial angle of star velocity respect to the initial one of S2 ($j=0  \rightarrow \varphi =\pi/2$) counted counterclockwise. Note that we assume S2 as starting from its apocentre so that given $x$ as the axis joining the  star with the SMBH, its initial velocity is along $y$, assuming $\dot{y}_0>0$ to give a counterclockwise revolution around the origin which is assumed in the system c.o.m.. These $x,y$ axes are on the motion plane (in Schwarzschild's geodesics the motion is planar). 
The mass of S2 has been estimated $m_*=13.6$ M$_\odot$ \citep{habibi2017}, that in our units is $m_*=3.192\times 10^{-6}$ and its initial conditions are 

$$\mathbf{r}_{0*} =(0.942762, 0, 0), 
~\mathbf{v}_{0*} = (0, 0.349792, 0).$$

Note that the center of mass of the Sgr A$^*$--S2 pair is within the SMBH Schwarschild's radius, being $r_{c.o.m.}\sim 0.074 r_S$, so any movement of the SMBH around its c.o.m. is by definition unobservable. \\
It is relevant noting that the stars of Tab.  2 are all characterized by pericenter distances greater than the tidal-disruption radius of Sgr A*, that is the distance to the SMBH within which an approaching star would be tidally disrupted, which is roughly given by: 
\begin{equation}
r_t \approx R_{*} \left( \frac{m_\bullet}{m_{*}} \right)^{1/3}.
\end{equation}

Actually, assuming that all the hypothetical stars considered (with $j \geq 1$) have the same mass and radius of S2 as given in \cite{habibi2017}, namely $m_{*}=13.6$ M$_\odot$ and $R_{*}=5.53$ R$_\odot$, we find that
$r_t \simeq 21 r_S $, and the innermost star ($j=9$ in Tab. 2) has a pericenter distance to Sgr A* of $r_p \sim 33 r_S \simeq 1.6 r_t$.
\\

For all the stars of Tab. 2 we computed the ``exact'' (Eq. \ref{eq:angshift}) Schwarzschild apsidal line precession per orbit, $\delta \varphi$. Figures 1 and 2 show, respectively, $\delta \varphi$  versus pericenter distance for the cases of Tab. 2
(excluding, for the sake of display, the innermost plunging $j=9$ star) and the dependence of $\delta \varphi$ on the mass of the central BH for the $j=0$ star. 
It is evident from Fig. 1 and Tab. 3 how the angular precession significantly grows for pericenters below $0.1$ mpc ($\sim 0.17$ that of S2), reaching up to $9$ degrees for the innermost plunging, quasi radial, orbit ($j=9$). The $\delta \varphi \propto m_{BH}^2$ behavior in Fig. 2 reflects the dominant $1$PN contribution (in this regime of c.i.) to the angular precession, which actually scales as $m_{BH}^2$. 

Table 3 compares the values of the angular shift per radial period of the set of cases in Tab.  2 as obtained by computation of the ``exact'' integral formula in Eq. \ref{eq:angshift} and those by numerical orbit integrations considering PN terms up to 3PN. 
The values in Tab.  3 indicate that representation of the actual $\delta \varphi$ in a Schwarzschild's geodesic via the 1PN terms, only, progressively underestimates the correct one for larger eccentricities (smaller pericenters) leading to a relative error $-2.1\%$ for the innermost plunging case of Tab.  2. 

The best approximation of the exact $\delta \varphi$ with the inclusion of only 1PN terms is of course obtained in the $j=0$ case where it gives 3 exact digits. 
Note that for inner orbits the 1PN term gives an angular precession which is up to $\sim 660$ arcsec smaller than the exact one, while the inclusion of the 2nd and 3rd order PN terms reduces this discrepancy to $\sim 0.5$ arcsec.  Figure 3 displays this fact by showing the relative difference (in absolute value) between the 'exact' $\delta \varphi$ and that evaluated by direct orbit integration using a 1PN and a 3PN approximation (black curves).

So the conclusion is that while for S2 a 1PN approx gives a quite accurate estimate of the apsidal line precession of the orbit, this accuracy gets rapidly worse for more penetrating S-stars, so that the inclusion of the 2PN term is important to have an accurate estimate of the precession of the orbit and in order to possibly test GR in the strong-field regime. Note, additionally, that from Tab.  3 it can be noticed that there isn’t a significant improvement between using a 2PN and a 3PN approximation for the cases considered here, so that a 2PN approximation is sufficient. \\

For the sake of completeness, since orbits with decreasing pericenter distances can be obtained not only decreasing the orbital angular momentum at fixed energy but also decreasing the orbital energy at fixed angular momentum, we also considered a second set of stars with the same $\mathbf{L}$ of S2 and reduced $E$, that corresponds to a progressive decrease of the semi-major axis and of the orbital period. In this case all the stars have the same eccentricity of S2, namely  $e=0.884649$.
In Tab. 4 we give the initial conditions for this set of orbits, with each star starting from its apocenter in the $x$ axis with initial velocity along the positive  $y$ axis. The star with $j=0$ is again S2, and stars with index $j>0$ have the same pericenter distances than stars with same $j$ in Tab. 2. 

We compared also for this new set the ''exact'' value of the apsidal line precession $\delta \varphi$ with the value obtained through numerical integration with our one-body code using a PN approximation up to 3PN order, as reported in Tab. 6 and plotted in Fig. 3 (red curves). In this case the numerical integration with PN terms gives a worse estimate of the exact $\delta \varphi$, comparing stars with same index $j$ (excluding $j=0$) of Tab. 2 and Tab. 4, as it is evident looking at Fig. 3.




\begin{figure}
\label{Fig:1}
\centering
\includegraphics[width=0.4\textwidth]{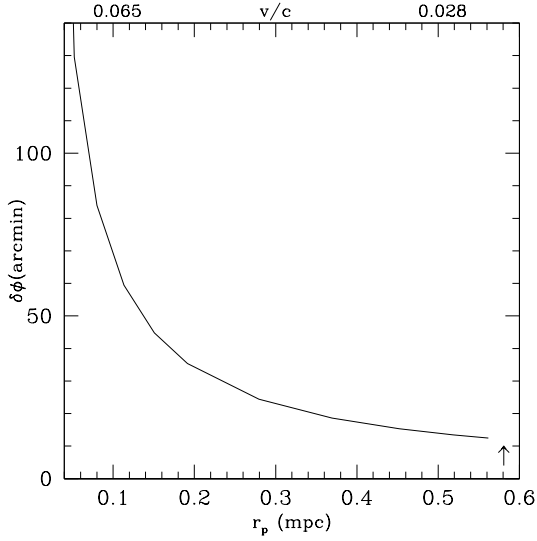}
\caption{Schwarzschild ``exact'' angular precession (in arcmin) per radial period versus pericenter distance (see Tab. 3). The vertical arrow refers to the star  S2 ($j=0$ in Tab. 2 and 3).
}
\end{figure}

\begin{figure}
\label{Fig:2}
\centering

           \includegraphics[width=0.4\textwidth] {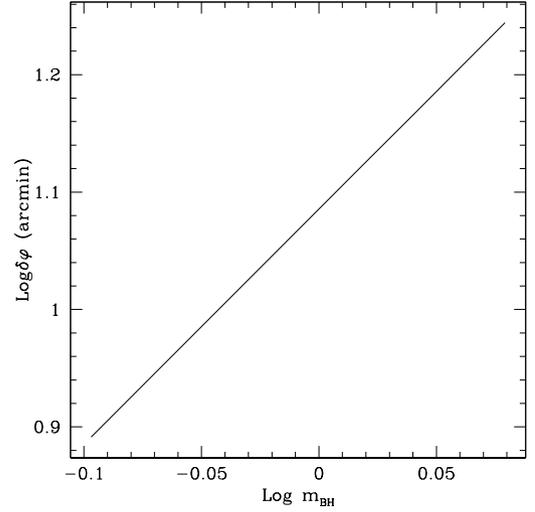}
               \caption{Schwarzschild logarithmic ``exact'' angular precession (in arcmin) per radial period versus central BH mass (in units of the assumed $m_\bullet$) for the $j=0$ star of Tab. 2.}
               
\end{figure}



\begin{figure}
\label{Fig:3}
\centering
\includegraphics[width=0.5\textwidth]{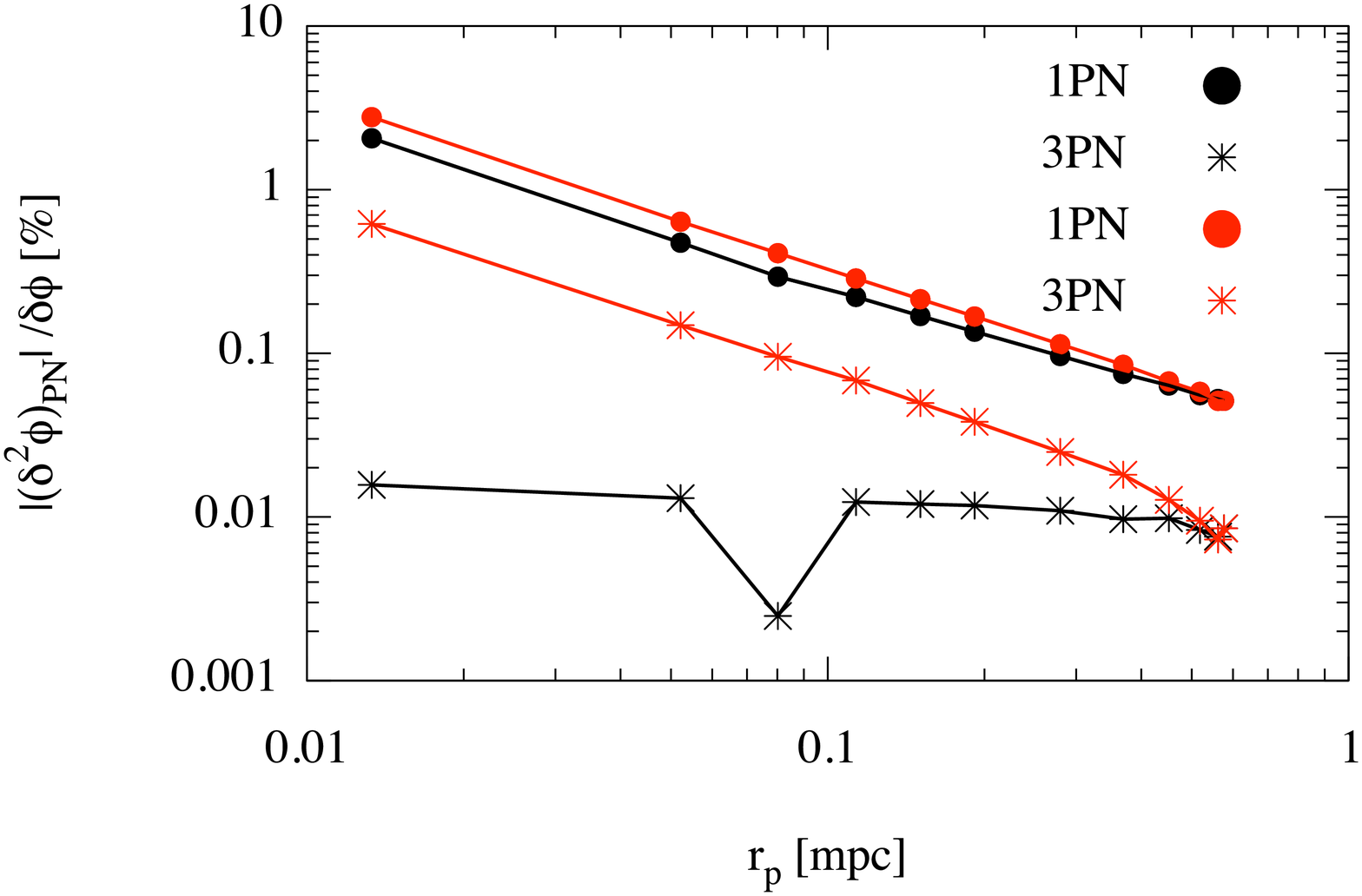}
\caption{For the cases of Tab.  2 (in black) and of Tab.  4 (in red): relative (percentage) difference in absolute value between the ``exact'' precession and that evaluated via direct integration of the orbital motion with a PN approximation up to 1PN and up to 3PN order, namely $|\delta^2 \varphi|/ \delta \varphi$ in function of pericenter distance.
}
\end{figure}

\subsubsection{A note on the 'observability' of apsidal precession}
 Let us remember here that GRAVITY is an interferometer that combines the light of the four VLT telescopes at Paranal Observatory in Chile (ESO), making it the most accurate instrument available for astrometric measurements \cite{abu17}. What is actually observable by experiments like GRAVITY it is not $\delta \varphi$ but rather the projected angular separation $\Delta \psi$ at two consecutive passages of an S-star at its pericenter (or apocenter).
Assuming as known the distance of the S-star cluster, we can easily compute the $\Delta \psi$ of stars of Tab.  2 under the (optimal) hypothesis they move on a plane orthogonal to the line of view (see following sect.  3.1 and Fig.5).
In particular, we calculated $\Delta \psi$ at the apocenter, where it is maximum. Taking into consideration that the astrometric accuracy of the GRAVITY instrument is between $10$ and $100 \, \mu as$ \citep{abu17}, we checked what is the maximum pericenter distance down of which the difference in $\Delta \psi$ evaluated considering once 1PN and once 2PN approximation in orbit integration becomes observable upon this accuracy, assumed to be equal to 100 $\mu as$, to be conservative. 
We found that a distinction between 1PN and 2PN case would be observable only for stars with pericenter distances smaller than $0.0349028 \, mpc$ (about $0.06$ that of S2).

\subsection{About the quality of apsidal line shift `analytical' approximations}
Various expressions have been proposed in the literature to evaluate the relativistic angular precession per orbit at orders above 1.
We cite at this regard \citet{dam88,oht89,kop94,kop20,ior20a,ior21,tuc19}.
There are many sophisticated considerations to do and this leads to some discrepancies among the results obtained by various authors.
In spite of an initial controversy among 2PN results obtained in
\cite{kop94} and \cite{ior20a}, solved in a correction expressed in \cite{ior21}, these authors agree in providing a result for the 2PN contribution to the angular precession coherent with that initially found by \cite{dam88}.

Taking into account results of the mentioned papers and the discussions contained therein, for the scopes of our present work we adopted the expression in Eq. 3.12 of \cite{dam88}, originally expressed in terms of specific orbital energy and angular momentum and that we here write in terms of the semi-major axis $a$ and the eccentricity $e$ of the orbit:

\begin{equation}
\begin{split}
\tilde{\delta} \varphi
=& \frac{6 \pi G m_\bullet}{c^2}\frac{1}{a(1-e^2)} +\\
+& \frac{\pi G^2 m_\bullet^2}{2 a^2 c^4} \Biggl[ \frac{3}{(1-e^2)}(-5+2 \eta)+ \frac{15}{(1-e^2)^2} (7-2 \eta) \Biggr] + \\
+&\frac{\pi G^3 m_\bullet^3}{64 a^3 c^6}  \Biggl[ \frac{24}{(1-e^2)}(5-5 \eta +4 \eta^2)+\\
+&\frac{1}{(1-e^2)^2} (-10080+(13952-123 \pi^2) \eta -1440 \eta^2) +\\
+&\frac{5}{(1-e^2)^3}(7392+(-8000+123 \pi^2) \eta +336 \eta^2) \Biggr],
\end{split}
\label{blanchet_eta}
\end{equation}

which, for negligible test particle mass (limit $\eta \rightarrow 0$), simplifies in

\begin{equation}
\begin{split}
\tilde{\delta} \varphi =& \frac{6 \pi G }{c^2}\frac{m_\bullet}{a(1-e^2)} + 
\frac{3 \pi G^2}{2c^4} \frac{m_\bullet^2 (30+5e^2)}{a^2(1-e^2)^2} + \\
+& \frac{3 \pi G^3}{2c^6} \frac{m_\bullet^3 (280+105e^2+\frac{5}{4}(1-e^2)^2)}{a^3(1-e^2)^3}.
\end{split}
\label{blanchet_eta0}
\end{equation}

This agrees at O$(1/c^4)$ with Eq. 9 in \cite{tuc19}, which gives the precession up to 3PN order as

\begin{equation}
\begin{split}
\tilde{\delta} \varphi =& \frac{6 \pi G }{c^2}\frac{m_\bullet}{a(1-e^2)} + 
\frac{3 \pi G^2}{2c^4} \frac{m_\bullet^2 (30+5e^2)}{a^2(1-e^2)^2} + \\
+& \frac{3 \pi G^3}{2c^6} \frac{m_\bullet^3 (280+105e^2)}{a^3(1-e^2)^3},
\end{split}
\end{equation}





that differs from Eq. \ref{blanchet_eta0} for the term of order $c^{-6}$, $\frac{3 \pi G^3}{2c^6} \frac{\frac{5}{4} m_\bullet^3 }{a^3(1-e^2)}$.
\\

Using Eq. \ref{blanchet_eta0} we computed for the set of initial conditions of Tab.  2 the differences between the “exact” precession and this approximated estimate evaluated at 1PN, 2PN and 3PN order. The results are given in Tab.  5. \\ 
Comparing the results in Tab 3 and 5 the first thing to notice is that, using a direct numerical integration of the orbital motion with 0PN and 1PN terms only, the estimate of the angular precession per orbit is much more accurate than using the standard 1PN approximation $( \tilde{\delta} \varphi)_{1PN} = \frac{6 \pi G }{c^2}\frac{m_\bullet}{a(1-e^2)}$. In fact, an approximation of the actual $\delta \varphi$ via this $(\tilde{\delta} \varphi)_{1PN}$ leads to a relative error up to about $-10\%$ for the innermost plunging case of Tab.  2, while using the numerical integration it is about $-2 \%$.  \\
When considering also the 2PN correction, the approximated expression given by Eq. \ref{blanchet_eta0} gives a better estimate of $\delta \varphi $ for the stars corresponding to $j\leq 7$, compared to what obtained through numerical integration including the 2PN acceleration term. On the contrary, going to smaller pericenter distances the difference between the ``exact'' $\delta \varphi$ and $(\delta \varphi)_{2PN}$ grows more rapidly using the approximated expression of Eq.\ref{blanchet_eta0} than the numerical integration, and, in particular for the innermost star ($j=9$), the numerical integration gives a much better result. \\
Using the whole Eq.\ref{blanchet_eta0}, it gives a moderately better estimate of the ``exact'' $\delta \varphi$ for all the stars but for the innermost star ($j=9$), for which again the numerical integration gives a much better result. In any case, the complete Eq.\ref{blanchet_eta0} gives a good estimate of the ''exact'' $\delta \varphi$, leading to a relative error that is always smaller than $0.1 \%$ in absolute value. 

The dependence of the results on the finite test star mass can be studied using the expression given by Eq. \ref{blanchet_eta} in its dependence on $\eta$.
In the previous computations we have assumed $\eta=0$ because the estimated mass range for the S-stars \citep{habibi2017, habibi19} is $0.5 \lessapprox m_i/M_\odot \lessapprox 20$, that means a range $1.19\times 10^{-7}\leq \eta \leq 4.76\times 10^{-6}$, which we considered sufficiently low to allow $\eta=0$ in the various expressions used. However, it is proper a check of the validity of this assumption by Eq. \ref{blanchet_eta} by a simple comparison of its results with $1.19 \times 10^{-7} \leq \eta \leq 4.76\times 10^{-6}$ with those obtained letting $\eta=0$. The result is that, accounting for the finite, non zero, mass of the star, the precession slightly decreases at increasing $\eta$ varying, anyway, of less than $\sim 10^{-5} \, \%$. Therefore, the role of the mass of the star on the orbital precession is really negligible.

\section{The role of the SMBH spin}
The apsidal line precession caused by relativistic effects in addition to classical precession as caused by terms other than the gravitational monopole does not limit to the previously discussed Schwarzschild (no spin) contribution. 

A Kerr black hole is characterized by a spin vector that we define as ${\bf S}_\bullet=(Gm_\bullet^2/c)\chi \bf{s_\bullet}$, with the Kerr parameter $0\leq \chi \leq 1$ and ${\bf s}_\bullet$ unit vector. It determines a non spherically symmetric space-time, and the motion in such space-time depends on both the magnitude of the BH spin and its orientation respect to the orbital angular momentum of the moving object. For the scopes of this paper we work in PN approximation to account for the SMBH spin on a supposedly spinless star moving in its field.

The leading order in $\chi$ is at 1.5PN order and is the spin-orbit (SO) term \citep{bark75}, that writes as 

\begin{equation}
\begin{split}
& \mathbf{f}_{SO} = \chi \frac{G^2(m_\bullet +m)^2}{c^3r^3}\left( 
\frac{1+ \sqrt{1-4\eta}}{4}\right) \\
& \big\{
\left[12\mathbf{s}_\bullet\cdot(\mathbf{n}\times \mathbf{v})\right]\mathbf{n} 
 + \left[\left(9+3\sqrt{1-4\eta}\right)\dot r\right] \\
 &  (\mathbf{n}\times\mathbf{s}_\bullet) - \left[7+\sqrt{1-4\eta}\right](\mathbf{v}\times\mathbf{s}_\bullet)
\big\},
\end{split}
\label{eq:spinorbit}
\end{equation}

where $m$ is the mass of the generic $i$th star orbiting the BH, $\mathbf{n}\equiv \mathbf{r}/r$ is the unit vector pointing from the BH to the moving particle, $\mathbf{v}=\dot{\mathbf{r}}$, $\dot{r}=\mathbf{v}\cdot \mathbf{n}$, and, finally, $\cdot$ and $\times$ represent the usual scalar and vector product between vectors.

The SO term corresponds to the well known Einstein-Thirring-Lense effect which is a rotational frame-dragging effect, and so it acts centrifugally respect to gravity, reducing,  when $\mathbf{s}_\bullet \cdot \mathbf{L}>0$, the apsidal precession respect to the zero-spin Schwarzschild case, while when $\mathbf{s}_\bullet \cdot \mathbf{L}<0$ it adds to centripetal action of gravity, increasing the apsidal precession respect to the zero-spin Schwarzschild case. SO does not only contribute to the in-plane precession of the orbit, but it gives also a precession of the orbital angular momentum vector $\mathbf{L}$ around the BH spin axis.

The quadratic in $\chi$, 2PN, quadrupole contribution is \citep{bark75}
\begin{equation}
\begin{split}
\mathbf{f}_{Q}= \chi^2 \frac{3}{2}\frac{G^3 m_\bullet^2(m_\bullet+m)}{c^4r^4}&\big\{\left[5(\mathbf{n}\cdot\mathbf{s}_\bullet)^2 - 1\right]\mathbf{n}
+ \\
&-2(\mathbf{n}\cdot\mathbf{s}_\bullet)\mathbf{s}_\bullet \big\}.
\end{split}
\label{eq:quadrupole}
\end{equation}

The quadrupole term $\mathbf{f}_{Q}$ depends implicitly on $\eta$ through the time evolution of $\mathbf{s}_\bullet$ which is governed by (see Eq. 7 in \cite{val10})

\begin{equation}
\frac{d \mathbf{s}_\bullet}{dt}=\mathbf{\Omega}\times\mathbf{s}_\bullet,
~
\mathbf{\Omega}=\left[\frac{G(m_\bullet+m)}{2c^2r^2} \right]\eta\left( \frac{7+\sqrt{1-4\eta}}{1+\sqrt{1-4\eta}}\right)(\mathbf{n}\times\mathbf{v}).
\end{equation}

\begin{figure}
\label{Fig:4}
\centering
 \includegraphics[width=0.5\textwidth] {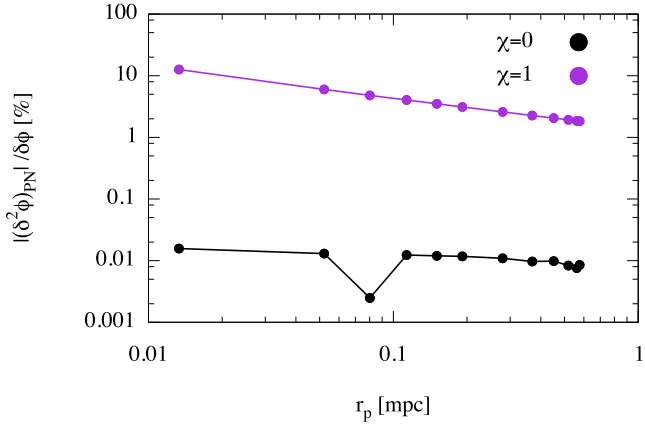}               
\caption{For the stars in Tab.  2: relative (percentage) difference in absolute value $|\delta^2\varphi| / \delta\varphi$ between the ``exact'' periapsis angular shift per radial period in the zero-spin case (Eq. \ref{eq:angshift}) and that obtained accounting for PN contribution up to 3rd order in the zero-spin case (in black) and in case of a corotating SMBH with maximal spin, $\chi=1$ (in magenta).}
\end{figure}

The quadrupole term gives an additional non-radial acceleration, which contributes both to the in-plane precession of the apsidal line and to the precession of the orbital plane. In the case of initial star velocity on the plane orthogonal to the black hole spin ($\mathbf{n}\cdot \mathbf{s}_\bullet=0$) it acts as an increase of the gravity intensity leading to a modest (due to $\chi^ 2/c^4$ dependence) addition to pericenter advance, whenever the pericenter distance is not too small ($f_Q \propto r^{-4}$). So, while the dependence of the spin-orbit effect on the orbit angular momentum is direct, for the quadrupole it is indirect (low values of $L$ means small pericenter distances and so an increased effect of the quadrupole acceleration).
\\

In this work we considered the simplified case in which the stars of Tab.  2 are on equatorial planar orbits, namely the case in which the spin of the SMBH is orthogonal to the plane of the orbit. In this case the spin gives only an additional contribution to the in-plane angular precession of the orbit and there is not a precession of the orbital plane.
In Tab. 7 we give the estimate of the difference between the expected orbital angular precession per radial period in the case of a zero-spin SMBH of mass $m_\bullet$ and that we numerically evaluated by orbital integrations with terms up to 3PN in the acceleration for the same SMBH mass spinning at maximal spin ($\chi = 1$) in the \textit{spin-up} co-rotating case ($\mathbf{L}\cdot \mathbf{s_\bullet}=|L_z|$), and \textit{spin-down}, counter-rotating case ($\mathbf{L}\cdot \mathbf{s_\bullet}=-|L_z|$). In Fig. 4 we plot the relative difference between the two values in function of pericenter distance in the co-rotating case.

Notice from Tab.  7 that the difference $\delta \varphi$ - $\delta \phi_{PN}$ is not equal in modulus in the co-rotating and counter-rotating cases. 
Actually, while the spin-orbit term is anti-symmetric for $\mathbf{s}_\bullet \rightarrow  -\mathbf{s}_\bullet$, so that the contribution it gives to the angular precession in the co-rotating and counter-rotating  cases is equal in modulus but opposite in sign, the quadrupole term is symmetric. Therefore, in both cases the effect of the quadrupole term is to slightly increase the angular precession and this explains why the difference $\delta \varphi$ - $\delta \phi_{PN}$ is not equal in absolute value in the co-rotating  and counter-rotating cases. 



\subsection{Measuring the spin of Sgr A* through the apsidal line shift}

Assuming a fixed value for the mass of Sgr A*, namely $m_\bullet =4.261\times 10^6$ M$_\odot$ \citep{abu20} ($m_\bullet=1$ in our units), and an equatorial planar orbit (SMBH spin orthogonal to the orbital plane), a measure of the in-plane precession of the pericenter would (in principle) allow to get the value of the spin of the SMBH. \\
However, this is not really true as we have to take into consideration what is the astrometric accuracy of the GRAVITY instrument \citep{abu17}.
A possible measure of the SMBH spin with the set of stars in Tab. 2 requires that the observed angular separation $\Delta \psi$  between the orbits in the Schwarzschild and Kerr cases (see Fig. 5) is over the threshold given by the astrometric accuracy of GRAVITY (assumed, to be conservative, as 100 $\mu as$). 
Since $\Delta \psi$ is maximum at the apocenter, we calculated for each star the angular separation $\Delta \psi$ between the two orbits at the apocenter (after one full radial period), for different values of the Kerr parameter $\chi$ in the [0,1] interval and $\mathbf{s}_\bullet = (0,0,1)$ (co-rotating case), assuming the ideal scenario in which all the stars are on face-on orbits, that is the case in which $\Delta \psi$ is the maximum possible. The extension of this study to the counter-rotating case is straightforward. \\
At present, we are not in the position of obtaining information about the spin of Sgr A* from the orbit of S2, because the angular separation $\Delta \psi$ after one full radial period between the orbit in case of a non-spinning SMBH and that in case of an SMBH even with maximal spin $\chi=1$ is $\Delta \psi= 15.203 \, \mu as $ which is lower than the astrometric accuracy of GRAVITY, conservatively assumed $100 \, \mu as$.\par
However, we could measure the spin if deeper S-stars, with smaller pericenter distances to Sgr A* than S2, were detected. In Fig. 6 is shown, for the set of deeper S-stars of Tab. 2, the minimum value of $\chi$ for which the orbit in the case of a non-spinning SMBH and in the case of a spinning SMBH with Kerr parameter $\chi$ can be distinguished within the astrometric accuracy of GRAVITY. The result is that for stars with a pericenter distance to the SMBH that is small enough (those corresponding to an index $j$ from $j=6.5$ to $j=9$, having  $r_p \lesssim r_{p_{S2}}/4$, $r_{p_{\rm{S2}}}$ being the S2 pericenter distance), the spin would, in principle, be measured.  Obviously, the smaller the  pericenter distance to the SMBH the smaller the threshold value of $\chi$ to pick a difference in the astrometric angle: for a hypothetical S-star with a pericenter distance $r_p \sim r_{p_{\rm{S2}}}/5$ 
the minimum value of $\chi$ that could be detected would be $\chi_{\rm{min}} \sim 0.6$, if $r_p \sim r_{p_{\rm{S2}}}/10 \, $ then $\chi_{\rm{min}} \sim 0.2$, and $\chi_{\rm{min}} $ is almost 0 for $r_p \sim r_{p_{\rm{S2}}}/100 \, $.


The $\chi_{\rm{min}}$ value has been computed considering one full radial oscillation of the stars in Tab. 2 which have all the same radial period ($T_r \sim 16$ yr). Therefore, in order to obtain an observational estimate of $\Delta \psi$, an observing time of at least $\tau_{\rm{obs}} = 16$ years would be necessary. 

Let us now consider the second set of hypothetical S-stars of Tab. 4,  having all the same eccentricity of S2 but progressively smaller semi-major axis (and so smaller orbital energy and radial period). In the assumed fixed observing time of $\tau_{obs} = 16$ yr as before, these new set of stars exploit several revolutions around the SMBH, so that the cumulative effect of the spin on the precession of the orbits results into a higher angular separation $\Delta \psi$ between the orbit resulting from a zero-spin SMBH and that of a spinning SMBH with Kerr parameter $\chi$. In fact, starting from the apocenter, we calculate now $\Delta \psi$ at the last apocenter reached  within 16 years, so namely after a number of revolutions $n= \rm{int} \left( \tau_{obs}/T_r \right)$, rounding down to the nearest integer. 
Also for this set of orbits we calculated the minimum value of the Kerr parameter $\chi_{\rm{min}}$ that could be detected in the accuracy limit, which is plotted in Fig. \ref{Fig:6} in function of the pericenter distance normalized to that of S2. \\
This Figure makes evident that, over an observing time of 16 yr, the possibility to distinguish the effect of smaller values of the Kerr parameter $\chi$ on the precession of the orbits with the second set of shorter period S-stars is significantly enhanced. Therefore, in order to measure the spin of Sgr A* through the precession of the orbits of S-stars, it would be better to detect a deeper S-star with the same eccentricity of S2 and smaller semi-major axis, and so shorter radial period, rather than stars with same semi-major axis $a$ of S2 and higher eccentricity $e$ yielding to same pericenter distance.

At the light of the above considerations, it comes out that the stars of Tab.  4  with shorter radial periods (those with $7\leq j \leq 9$) would still allow to measure the spin even over an observing time span of $\tau_{obs}=5$ yr. In Fig.  \ref{Fig:7} we plot the values of $\chi_{min}$ obtained in function of pericenter distance. It results that, in order to  measure the spin of Sgr A* in a time $\tau_{obs}= 5$ yr via the motion of S-stars close to the SMBH with same eccentricity of S$2$ but smaller semi-major axis, we would need to detect S-stars with pericenter distances $r_p$ smaller than $\sim r_{p_{S2}}/5\,$ and radial period $T_r$ smaller than $\sim$ 2 yr. Moreover, a star with $r_p$ smaller than $\sim r_{p_{S2}}/10 \,$ and radial period $T_r$ smaller than $\sim$ 0.5 yr, would allow to measure values of the spin higher than  $\chi_{min}=0.2$, while a star with $r_p$ smaller than $\sim r_{p_{S2}}/50$ and $T_r$ smaller than $\sim$ 0.05 yr would be sensitive to almost every possible value of $\chi>0$.\\

We remind that these results are obtained in the case of face-on orbits and of a SMBH with spin orthogonal to the plane of the orbit, that is the `optimal' scenario, because in this case the observed effect of the spin on the in-plane angular precession of the orbits is maximized. A further necessary step in this study will thus concern the variation of the inclinations of the orbits and the variation of the angle between the spin and the orbital angular momentum, that is more complicated because the orbits will no longer be planar but rather show a precession of the orbital plane, which, if detected, would represent the 'smoking gun' of the presence of spin of the SMBH.\\
At this regard we note that in their pioneering work  \cite{merwill10} by mean of orbital integrations of test stars with a version of the chain-regularized code \arwv reached the conclusion that stellar orbits limited within $\sim 1$ mpc from the SMBH are needed in order to possibly measure the spin of Sgr A*. This result is however in a different frame from ours, in what these authors investigated the compared effect of the SMBH spin on the orbital precession and that (classical) due to the perturbations by  other stars and stellar remnants distributed around the SMBH, while in our work we focused on the role of different PN terms on the orbital precession and on the orbital characteristics of stars for which the effect of the spin on the precession would be large enough to be observed with the GRAVITY instrument at the VLT.

\begin{figure}
\label{Fig:5}
\centering
\includegraphics[width=0.5\textwidth]{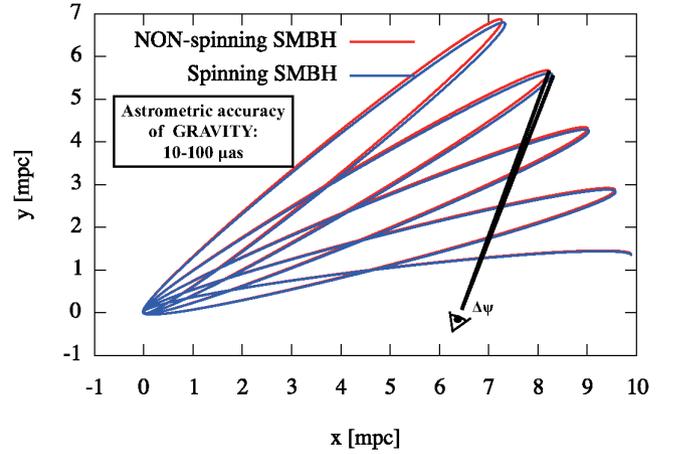}
\caption{Angular difference at apocenter, $\Delta \psi$, between the orbit in case of a non-spinning SMBH and in case of a spinning (co-rotating) SMBH. }
\end{figure}



\begin{figure}
\label{Fig:6}
\centering
\includegraphics[width=0.5\textwidth]{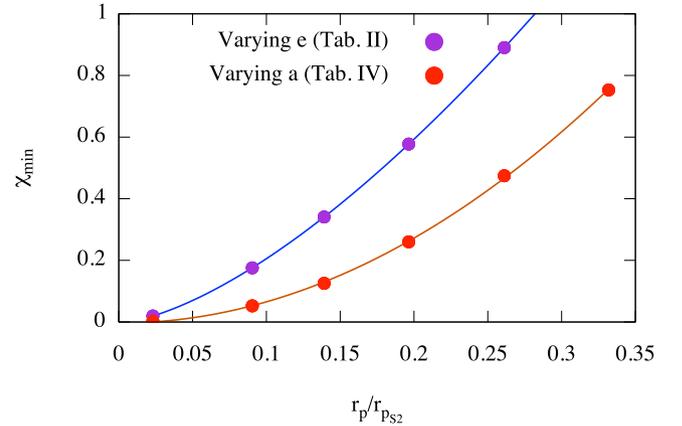}
\caption{For the two sets of deep S-stars of Tab.  2 (magenta points) and Tab.  4 (red points), and considering an observing time of $\tau_{obs}= 16$ yr, the figure gives the minimum value $\chi_{min}$ of $\chi$ for which the orbit in case of a non-spinning and in case of a spinning SMBH can be distinguished within the astrometric accuracy of GRAVITY (assumed 100 $\mu as$), in function of pericenter distance in units of the pericenter distance of S2.}
\end{figure}

\begin{figure}
\label{Fig:7}
\centering
\includegraphics[width=0.5\textwidth]{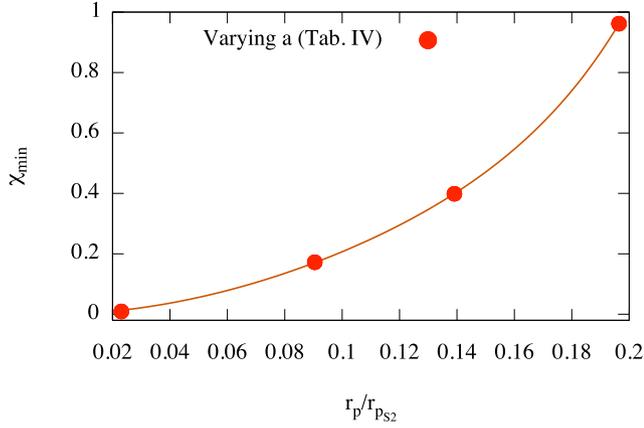}
\caption{As in Fig. 6 but for the set of deeper S-stars of Tab.  4 and an observing time of $\tau_{obs}=5$ years.}
\end{figure}

\subsection{Measuring the spin of Sgr A* through the gravitational redshift}
\label{Sec:III.B}
The effect of the spin of Sgr A* on the motion of S-stars can in principle be observed also through its role in the gravitational redshift.\\
General Relativity predicts that if a light source is in the vicinity of a massive body, the radiation is redshifted when detected far away from the body. The relation between the emitted and the observed frequency is

\begin{equation}
    \frac{\nu_{em}}{\nu_{obs}}= \sqrt{\frac{g_{00} (x^{\mu}_{obs})}{g_{00} (x^{\mu}_{em})}},
\end{equation}
where $g_{00}(x^{\mu})$ is the time component of the metric tensor, and $x^\mu$ are the space-time coordinates. Therefore the gravitational redshift is

\begin{equation}
    z_{G}=\frac{\nu_{em}}{\nu_{obs}}-1=  \sqrt{\frac{g_{00} (x^{\mu}_{obs})}{g_{00} (x^{\mu}_{em})}}-1.
\end{equation}
 For the derivation of this expression see for example \cite{ferrari2020general}.

Assuming that the space-time is described by the Schwarzschild metric \citep{sch16} it follows that 


\begin{equation}
    z_{G,S}= \sqrt{\frac{g_{00} (x^{\mu}_{obs})}{g_{00} (x^{\mu}_{em})}}-1= \sqrt{\frac{1-\frac{r_S}{r_{obs}}}{1-\frac{r_S}{r_{em}}}}-1 \approx \frac{1}{2}\frac{r_S}{r_{em}},
\end{equation}

the rightmost expression being valid in the limits $r_{obs}\gg r_{em}\gg r_S$.



Let us consider now the case in which the spacetime is described is described by the Kerr metric \citep{kerr1963}, such that
\begin{equation}
    g_{00}=- \left( 1- \frac{r_S \, r}{r^2+(\frac{a}{c})^2 \rm{cos}^2 \theta} \right),
\end{equation}
where $\theta$ is the angle between the positive spin axis and the position vector $\bf{r}$, $a= Gm_\bullet \chi/c$ and $\chi$ is the usual Kerr parameter.\\

In the same limit $r_{obs}\gg r_{em} \gg r_S$ of Eq. 23, it results 
\begin{equation}
\begin{aligned}
    z_{G,K} &\approx \sqrt{\frac{g_{00} (x^{\mu}_{obs})}{g_{00} (x^{\mu}_{em})}} -1 \approx \sqrt{  \frac{1- \frac{r_S}{r_{obs}}+ \frac{r_S}{r_{obs}^3}(\frac{a}{c})^2 \rm{cos}^2 \theta}{1- \frac{r_S}{r_{em}}+ \frac{r_S}{r_{em}^3}(\frac{a}{c})^2 \rm{cos}^2 \theta}  } - 1 \\
    &\approx \frac{1}{\sqrt{1- \frac{r_S}{r_{em}}+ \frac{r_S}{r_{em}^3}(\frac{a}{c})^2 \rm{cos}^2 \theta}} -1 \\
    &\approx \frac{r_S}{2r_{em}} - \frac{r_S}{2 r_{em}^3} \left(\frac{a}{c} \right)^2 \rm{cos}^2 \theta.
    \end{aligned}
    \label{gr_red}
\end{equation}

\noindent
Consequently
\begin{equation}
    z_{G,K} \approx z_{G,S} - \frac{1}{2c^2}\frac{a^2r_S}{r_{em}^3}  
    \cos ^2 \theta,
\end{equation}
and thus the gravitational redshift in the Kerr case is always smaller than in the Schwarzschild case, except for equatorial orbits ($\theta= \pi /2, 3 \pi /2$) where     $z_{G,K} = z_{G,S}=r_S/(2r_{em})$.

\begin{figure}
 \label{Fig:8}
\centering
\includegraphics[width=0.5\textwidth]{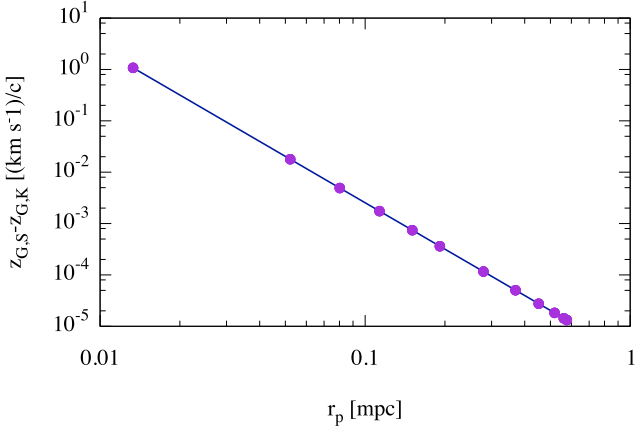}
\caption{Maximum difference between the gravitational redshift in case of a Schwarzschild SMBH and a Kerr SMBH in function of pericenter distance.}
\end{figure}

We want to see now whether the difference between the gravitational redshift in case of a Schwarzschild SMBH and a Kerr SMBH as evaluated at pericenter ($r_{em}=r_p$) where this difference is the highest, is observable with the spectroscopic accuracy of the instruments currently available. In order to do that, we consider the case in which the effect of the spin on the gravitational redshift is the maximum possible, namely we consider a SMBH spinning with maximal spin $|\chi|=1$ and test orbits such that at pericenter $\rm{cos} \theta=1$. In this case

 \begin{equation}
     z_{G,S} - z_{G,K} \approx  \frac{1}{8}\left(\frac{r_S}{r_p}\right)^3,
 \end{equation}
 
that corresponds to a relative deviation

\begin{equation}
\frac{z_{G,S} - z_{G,K}}{z_{G,S}}\approx z_{G,S}^2,
\end{equation}

which is quadratically small for all reasonable cases.
In Fig. \ref{Fig:8} we plot the difference of Eq. 27 in function of the pericenter distance, considering orbits with pericenter distances of the stars in Tabs. 2 and 4. This difference is extremely small for S2 (rightmost dot in the figure), $\sim 10^{-5} \, \frac{km \, s^{-1}}{c}$, and becomes slightly higher than $1 \, \frac{km \, s^{-1}}{c}$ only for the innermost star ($j=9$), that has a pericenter distance about $1/43$ that of S2. \\ 
In \cite{abu18}, the group led by R. Genzel at the MPE used the SINFONI spectrograph to measure the combined transverse Doppler effect and gravitational redshift of S2 at pericenter. Recently a new spectroscopic instrument has been mounted on the VLT UT4 telescope, named ERIS (Enhanced Resolution Imager and Spectrograph), that will have the same accuracy of SINFONI in the redshift measurement, up to $\sim 7 \, \frac{km \, s^{-1}}{c}$ \citep{gr2019}.  Unfortunately, this accuracy is not good enough to appreciate the difference  $z_{G,S} - z_{G,K}$ even for the innermost star ($j=9$) we considered and a maximally spinning, $\chi=1$, SMBH. \\
 So, the effect of the spin of Sgr A* on the gravitational redshift of potentially observable, deep S-stars, is too small to be detected with the current spectroscopic accuracy of ERIS. The future ELT (Extremely Large Telescope) will potentially allow to reach an accuracy of $\sim 1 \frac{km \, s^{-1}}{c}$ \citep{ESOelt}, but this would not change things significantly for this discussion, as the difference $z_{G_{Schw}} - z_{G_{Kerr}}$ becomes higher than $ 1 \frac{km \, s^{-1}}{c}$ only for the innermost star and only for a SMBH with maximal spin $|\chi|=1$. \\

\bigskip
\subsection{Mass-spin degeneracy}
It is crucial stressing that, if an independent measure of the mass of the SMBH is not available, there is clearly a degeneracy between the mass and spin of the SMBH for what regards the quantity of apsidal line precession.
Actually, as we previously said, in the co-rotating case the effect of the spin of the central SMBH corresponds to a decrease in the value of the angular precession of the orbit like it happens when the SMBH mass is reduced. In the counter-rotating case the effect is the opposite (increment of precession) which can be mimicked by an increased SMBH mass.\\
To quantify the combined role of the SMBH mass and spin, here we studied which value of the mass $m_\bullet$ of the SMBH would lead to recover the value of the angular precession previously obtained for S2 in the non-rotating case under the assumption $m_{\bullet, ref} =4.261\times 10^6$ M$_\odot$ in a 3PN approximation, this time letting the SMBH spinning at $\chi =1$ in the co-rotating and counter-rotating cases.  We got: 
\begin{itemize}
\item $m_\bullet = 1.00930 \, m_{\bullet, ref}$ in the co-rotating case ${\bf s}_\bullet =(0,0,1)$;
\item $m_\bullet = 0.99101 \, m_{\bullet, ref}$ in the counter-rotating case ${\bf s}_\bullet =(0,0,-1)$.
\end{itemize}

Therefore the relative deviation from $m_{\bullet, ref}$ is 
\begin{itemize}
\item $\Delta m_1 = 0.00930$ in the co-rotating case;
\item $\Delta m_2 = 0.00899$ in the counter-rotating case.
\end{itemize}

This result is interesting because if we consider the uncertainty on the value of the mass of Sgr A* given in \citep{abu20}, namely $\sigma_{m_\bullet, ref}/m_{\bullet, ref}=  0.00282 \,$, the above deviation  from $m_{\bullet, ref}$  is $> 3\sigma_{m_\bullet, ref}$. \\ Therefore, the apsidal line precession of the orbit of S2 that we obtained in the 3PN approximation in case of a Schwarzschild SMBH with mass $m_{\bullet, ref}$ can be obtained also in case of a Kerr SMBH with maximal spin ($\chi=1$) and mass $m_{\bullet, ref} + \Delta m_1$ if co-rotating or $m_{\bullet, ref} - \Delta m_2$ if counter-rotating, with $\Delta m_2,  \, \Delta m_1  \, > \, 3 \sigma_{m_\bullet, ref}$.
\\

This means that the simple measure of the in-plane apsidal line precession of the orbit is not sufficient to constrain both the mass and the spin of the SMBH.
However, information on the mass and spin of Sgr A* is enclosed also in the gravitational redshift of the star (Eq. \ref{gr_red}).
Therefore, from a theoretical point of view, the degeneracy between the mass and the spin of the black hole could, in principle, be removed by measuring both the gravitational redshift of the star and the apsidal line precession of the orbit.\\

\subsection{Non planar orbits}
In case of a generic non-planar orbit around the spinning BH (angle between spin axis and orbital angular momentum different from 0, $\pi$/2,  $3\pi$/2, $\pi$), there is also a precession of the orbital plane and so a time variation of the inclination $i$ of the orbit and of the line of nodes $\Omega$ \citep{will08, merwill10}. This precession of the line of nodes would be, actually, a clear signature of a spinning black hole, in the assumption that the orbits showing the line of nodes precession are so close to the BH such to be not influenced by significant local  perturbations inducing a line of nodes precession of `classical' origin.

The parameters we wish to measure would be, then, four: mass of the BH, magnitude of the spin, and two angles to determine the direction of the spin. \\
For this reason, in order to answer this problem we would need to extend the study done in Section 3.1 varying the inclination of the orbits and the angle between the spin and the orbital angular momentum. For each configuration we would obtain, in addition to the in-plane apsidal line precession per radial period, the variation of $i$ and of  $\Omega$ per orbital period. The idea is that to have a set of simulations that would allow, if an S-star with a small enough pericenter distance were detected, to obtain the mass of the BH and the magnitude and the direction of its spin, starting from the measure of the in-plane precession of the apsidal line, of the gravitational redshift, of the variation of the inclination $i$ of the orbit and of the line of nodes $\Omega$.

\section{Conclusions}
Main conclusions of this paper can be summarized:
\\
\begin{itemize}
    \item in the Galactic central region, assuming spherical symmetry for the distribution of matter and a non spinning local SMBH, orbital precession is given by both an amount of classical retrograde motion of the apsidal line and a general relativistic prograde contribution, which is dominant in the innermost region; 
    \item the standard first order Schwarzschild-Einstein approximation of the  relativistic pericenter advance which accounts very well for the Mercury's classically unexplained 43 arcsec per century advance, is not enough to justify with high accuracy the pericenter precession of stars deeply orbiting in the strong field around a supermassive black hole (like Sgr A$^*$ in our Galaxy); 
    \item an improvement in this direction is given by high precision numerical integrations of deeply plunging star orbits in a post-Newtonian scheme up to 3rd order. We showed that an integration accounting for 1PN terms only leads, for our set of test orbits, to a relative error in the determination of the pericenter precession up to about $2 \%$ with respect to the 'exact' precession expected in a Schwarzschild's geodesic, while including the 2PN and 3PN terms the error is always lower than about $0.2 \%$, giving thus a much better estimate of the 'exact' precession;
     \item given the astrometric accuracy of GRAVITY, a distinction between the orbit in case of 1PN only and 1PN+2PN approximation would be observable only for stars with pericenters smaller than $0.0349028 \, mpc$ (about $0.06$ that of S2);
    \item the approximated `analytical' expressions to 3rd PN order for the estimate of the pericenter advance are compared with the direct orbital integrations and with the 'exact' determination of the advance in a Schwarzschild's geodesic, finding that they loose accuracy respect to the orbital integrations whenever the test star pericenter distance decreases enough;
    \item the smoking gun of a spinning SMBH in our Galaxy would be the evidence of precession of the line of nodes for S-stars with pericenter distance which is a fraction of that of S2;
    \item we showed how the better way to distinguish observationally the spin contribution to the no-spin pericenter advance would be the study of short period stars around the Sgr A$^*$ object, whose discovery is one of the aim of the future improvement of the GRAVITY experiment at the ESO Very Large Telescope facility. In particular, we showed that to measure the spin of Sgr A$^*$ in an observing time less than 5 years an S-star with a pericenter distance not greater than $1/5$ that of S2 would be needed;
    \item the degeneracy of the mass-spin contribution to the deeply plunging motion, even in the hypothesis of negligible orbital perturbation of the nearby stars, is hard to solve at the light of present observing capabilities;
    \item a potential way to solve the above mentioned degeneracy would be coupling redshift and precession information, but the spin-induced redshift contribution is too small to be detected at present state of art of spectroscopic observations.

\end{itemize}

\section*{Acknowledgements}
We thank Pau Amaro Seoane and Stefan Gillessen for their useful comments and suggestions along the development of this work. 
We also thank an anonymous referee for comments helful to improve the paper.

\section*{Data availability} 
The output data of the simulations of this paper are available upon request to the corresponding author. Their use is subjected to proper citation.




\bibliographystyle{mnras}
\bibliography{reference} 


\bsp	
\label{lastpage}
\end{document}